# Talbot photolithography optimization with engineered hybrid metal-dielectric mask: High-contrast and highly-uniform Talbot stripes

Yu.E. Geints[1,*], I.V. Minin[2], and O.V. Minin[2]

[1]V.E. Zuev Institute of Atmospheric Optics SB RAS, 1 Zuev square, 634021, Tomsk, Russia
[2]Tomsk State Polytechnic University, 36 Lenin Avenue, 634050, Tomsk, Russia.
*ygeints@iao.ru

**Abstract:**
Conventional projection Talbot lithography usually employs opaque (amplitude) or transparent (phase) masks for creating a periodic array of Fresnel diffraction fringes in the photosensitive substrate. For particular mask design the longitudinal periodicity of Talbot carpet can be avoided producing quasi uniform striped pattern (Talbot stripes). We propose a novel hybrid amplitude-phase mask which is engineered for obtaining extremely smooth Talbot stripes and simultaneously high lateral optical contrast and extreme spatial resolution better than a third of laser wavelength. By means of the numerical simulations, we demonstrate the robustness of produced striped diffraction patterns against mask design deviation and light incidence angle variations. The reproducibility of the Talbot stripes is reported also for 1D and 2D metal-dielectric projection masks.

**Keywords**: Talbot lithography, wavelength-scaled diffraction grating, optical contrast, projection mask.

## I. Introduction

Wavelength-scaled diffraction gratings (DGs) with a ridge period of the order of a working optical wavelength are capable to create a subwavelength periodic structure of an optical wave passed through or reflected from the grating [1]. In this case, due to the interference of waves of different diffraction orders (DOs), a periodic spatial pattern of alternating maxima and minima of field intensity appears behind the grating in the Fresnel zone, which is often referred in the literature to as the Talbot carpet in the name of H.F. Talbot, who presented the first experimental observation of periodic self-repeating images from periodic DG [2,3]. In particular, this stable structural light periodicity has found practical application in optical Talbot lithography (TL) for optical microstructuring of photosensitive substrates [4-8]. It should be noted that in the latter case, the prospects for using of wavelength-scaled DGs are directly related to their ability to give diffraction patterns with subwavelength spatial resolution [9].

In classical Talbot photolithography, one technical challenge that reduces the accuracy of photoprinting is the need for accurate positioning of the photosensitive matrix (a sample) in the area of the most sharp projection mask image, which usually does not exceed several optical wavelengths. For solving this problem, the displacement Talbot lithography (DTL) [10] can be used implying that the irradiated substrate, or projection mask is not fixed in space but cyclically moves in the range of several Talbot wavelengths ($z_T$), thus performing spatial integration of the Fresnel diffraction pattern in the longitudinal direction. This technique improves the depth-of-field of the mask images and also eliminates the need for physical contact between the mask and the sensitive sample, the surface of which may not be perfectly flat. To date, DTL technology has improved the speed and accuracy of photoprinting and fabricated subwavelength diffraction gratings [11-13], as well as very complex periodic photonic structures [14].

The need for mechanical movement of the photoresist imposes rather stringent requirements for the stability of the entire optical system, which can worsen the quality of DTL photoprinting. At the same time, from the viewpoint of the Fresnel diffraction theory the Talbot diffraction carpet results from the interference of optical waves scattered in different DOs, which provides high periodicity of the Talbot carpet in all spatial dimensions. Manipulating the optical power in the diffractive orders can significantly reshape the entire diffraction pattern and produce a mask image, e.g., in the form of equidistant stripes with low longitudinal intensity modulation. This can completely eliminate the problem with the high longitudinal variability of the self-images in the Talbot photolithography and the need of mechanical movement of the photosensitive substrate in the DTL.

One way to control the diffraction orders is proposed in [15] by employing an oblique illumination of optical phase mask with UV-laser. By selecting the specific illumination angle, the contribution of higher DOs can be significantly reduced and longitudinally uniform diffraction fringes can be obtained. This provides the photoprinting of periodic structures with high aspect-ratio (linear-to-lateral ~ 30) and subwavelength transverse resolution close to the diffraction limit (~ $\lambda/2$).

Another method of tailoring Talbot carpet is demonstrated in [16] by equipping a wavelength-scaled 1D projection transparence with a vignetting metal mask superimposed directly on the tops of the phase steps. With an appropriate choice of the grating pitch and width of the amplitude mask, a diffraction pattern consisting of longitudinal equidistant fringes with alternating brightness is also formed. Worthwhile noting that the very idea of combining the amplitude and phase masks to increase the spatial resolution of the near-field focusing of optical radiation based on anomalous anodization effect [17,18] for the diffraction grating was first proposed in [19]. A similar hybrid metal-dielectric mask was investigated in [20] for the purpose of increasing DTL contrast.

In this paper, inspired by the works [15, 16, 20] we investigate the potential of a hybrid amplitude-phase projection mask for the Talbot photolithography in terms of obtaining the contrasting but longitudinally homogeneous Talbot self-images. Using the finite-elements (FEM) and finite-difference time-domain (FDTD) numerical calculations of the near-field diffraction of a monochromatic light wave on a metal-dielectric wavelength-scaled diffraction grating, we show that under certain structural parameters of such a hybrid mask homogeneous diffraction fringes, hereafter referred as Talbot stripes, are produced with minimal longitudinal intensity variation and extremely high transverse optical contrast (> 30 dB). Talbot stripes can be formed because the metal part of the hybrid mask almost completely shields the direct radiation coming in the zero DO, thus providing the separation of optical waves propagating only in the spatial angles of the negative and positive first diffraction orders. By using elliptically shaped phase ridges instead of a rectangular phase mask, one can further increase the spatial resolution of the diffraction fringes up to the subwavelength values of ($\lambda$/3.6) which is beyond the Abbe diffraction limit and almost close to the value achieved in the Talbot immersion lithography [21]. Moreover, we demonstrate the universality of Talbot stripes generation also for the case of a hybrid 2D grating illumination having a wafer metal-dielectric structure.

## II. Computer model and simulation methodic

From now on, the formation of Talbot diffraction stripes is considered for several 1D projection masks shown in Figs. 1(a-c). Obviously, each mask is a binary phase DG with the steps producing the π-phase shift. Such DG is characterized by the pitch $d$, phase step height $h$, width $w$, and filling factor $\kappa = w/d$ (Fig. 1(d)). The refractive index (RI) of the phase grating material is designated as $n$. It is assumed that a phase DG is made of optical glass with zero light absorption and placed in air with $n_0 = 1$.

Two types of Talbot projection masks shown in Figs. 1(b) and 1(c) are additionally equipped with amplitude masks made in the form of metal strips of width $w_m$, thickness $h_m$ and opening ratio $\gamma = 1 - w_m/d$. The layout of metal masks on the phase DG corresponds to that in [16] and [20], respectively. For certainty, in the numerical simulations several model parameters are fixed, namely: $n = 1.5$, $h_m = 50$ nm.

The projection mask is illuminated by a linearly polarized monochromatic optical wave with the amplitude $E_0 = 1$ V/m and wavelength in the UV spectrum region $\lambda = 375$ nm, which is typical for commercial TL devices. A two-dimensional Talbot carpet is formed behind the mask in the Fresnel diffraction zone.

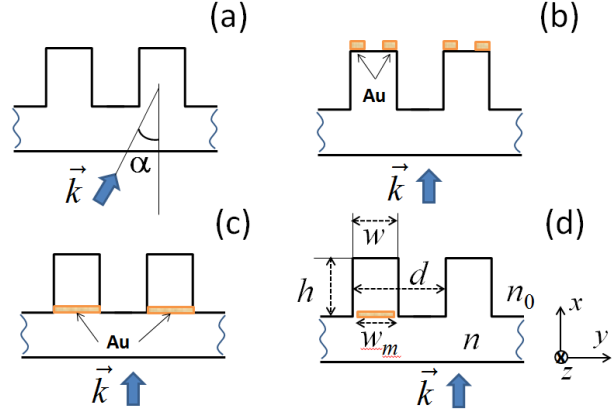

Fig. 1. (a-c) Projection masks for Talbot stripes formation: (a) Oblique-incidence TL with normal phase DG [15], (b) Amplitude-vignetting phase DG [16], and (c) Proposed hybrid phase DG with amplitude-screening; (d) Geometrical scheme of the hybrid metal-dielectric TL mask.

The optical wave diffraction on the TL masks is simulated based on the numerical solution to the Helmholtz equation for the electromagnetic field employing the finite element method implemented by the commercial software package COMSOL Multiphysics (version 5.1). 2D-*xy* geometry of the computational domain is used, whereas along the grating axis (*z*) the lattice is assumed to be infinite. Due to the periodicity of the DG in the *x*-direction, the computer model of the optical metastructure is constructed for only one grating period. On the lateral boundaries of the building block, the Floquet-Bloch periodicity boundary conditions are set. In the direction of light incidence the conditions of perfect field matching (PML) are used. The numerical accuracy of the solution is ensured by the adaptive computational mesh, which nodes are clustered in the regions of sharp dielectric permittivity gradients of the medium (metal ridges). The maximum step of the computational spatial grid is chosen as $\lambda/30$ inside the DG and $\lambda/15$ in the environment.

## III. Results and discussion

### a. Diffraction analysis of Talbot stripes

Examples of the two-dimensional Talbot stripes are shown in Figs. 2(a-c) for the case of UV-illumination of different TL masks depicted in Figs. 1(a-c). The corresponding longitudinal intensity distributions in the diffraction fringes are presented in Figs. 2(d-f). As seen, in all cases the classical Talbot diffraction pattern consisting of multiple mask images with the characteristic spatial Talbot scale $z_T = \lambda / \left(1 - \sqrt{1 - \lambda^2/d^2}\right)$ [22] is not observed. Instead, the masks produce a set of longitudinal stripes with different transverse widths and periodicity. Thus, for the two hybrid metal-dielectric masks in Figs. 1(d) and 1(c), the transverse spatial frequency of the diffraction fringes is twice as high as for the oblique-illuminated ordinal phase mask in Fig. 1(a). Note that in the latter case according to the data of [15], the wavelength of a laser source is slightly shorter, $\lambda = 360$ nm, and the height of the phase step is lower, $h = 200$ nm. Thus, it is clearly visible that in the proposed variant in Fig. 2(c), the Talbot diffraction stripes possess the smallest half-width and as it will be shown below, the best transverse optical contrast.

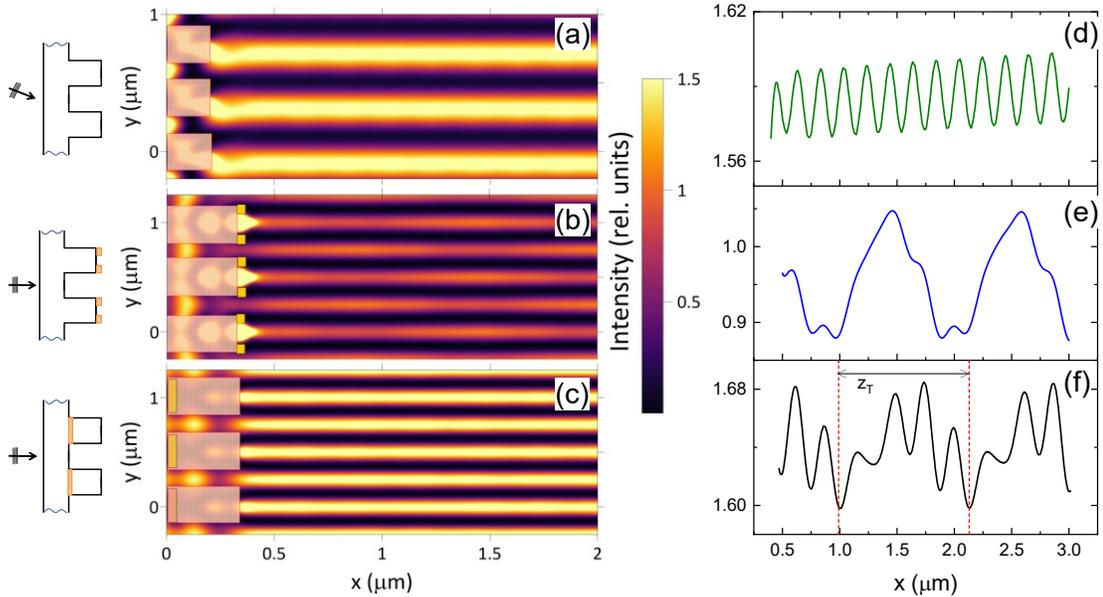

Fig. 2. (a-c) Examples of Talbot stripes obtained with different projection schemes; (d-f) Talbot stripe longitudinal intensity profiles.

Worth noting, the disappearance of the longitudinal periodicity of mask self-images is obtained only for certain geometry of the DG ridges and the angle of the illumination. In all other cases the well-known high-frequency Fresnel phase mask self-images are realized [3].

Qualitatively, the formation of Talbot stripes can be analyzed in terms of optical wave diffraction on a periodic lattice. Such diffraction element creates a sequence of diffraction maxima in the environment due to the interference of optical waves scattered in different DOs. The deflection angle $\theta_m$ of $m$-order diffracted wave from the direction of incidence increases as its order $m$ increases according to the following formula: $\sin\theta_m = \pm m\lambda/d + \sin\theta_i$, where $\theta_i$ is the angle of grating illumination. The arising of longitudinal stripes in the Talbot diffraction carpet corresponds to a situation where all but a few orders destructively interfere beyond the diffraction grating. For this purpose one can choose such an angle of mask illumination $\theta_i$ that, e.g., all DOs with $m > 2$ will have a deflection angle $\theta_m > \pi/2$ [15]. Then, only zero and one of the first DOs will persist in the scattered field, and by the interference will result in the desired striped Talbot carpet.

Another way of the eliminating the longitudinal periodicity of the diffraction fringes considered in the present work is the spatial filtering the diffraction orders of the optical field using an opaque amplitude mask. Due to the peculiarities of wave diffraction on a phase step such a hybrid amplitude-phase mask will effectively block the zero diffraction order. Provided the DG ruling is small compared to the wavelength, $d \leq 2\lambda$, only two first DOs with $m = \pm 1$ will remain in the consideration.

Note that the observed transformations of the diffraction patterns cannot be explained by the classical Talbot theory even for a simple rectangular phase DG [23]. The reason is that the analytical expressions for spatial filed distribution are derived usually in the paraxial approximation of the Rayleigh-Sommerfeld diffraction integral [24], which is valid only under the condition $d \gg \lambda$. In the case considered here, the diffraction grating has a spatial scale of phase relief comparable with the wavelength of the illuminating light, $d \sim \lambda$. This cardinally changes the diffraction patterns on the wavelength-scaled DG and leads to the redistribution of the optical field energy into higher-frequency spatial harmonics. Therefore, the full-wave calculations of the electromagnetic field components are more appropriate for simulation of the near-field diffraction on a mesoscale projection mask and quantifying the structural parameters of Talbot diffraction patterns.

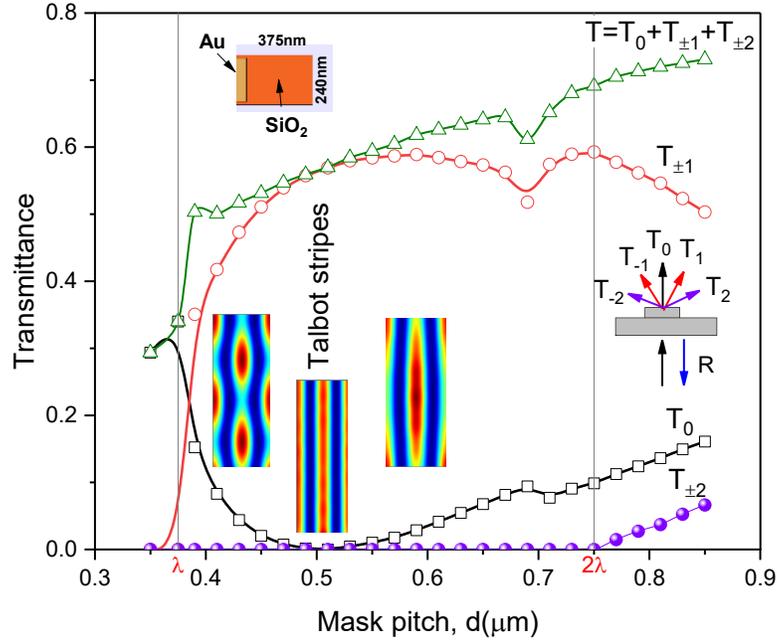

Fig. 3. Optical transmittance of different DOs for the proposed hybrid mask (Fig. 1(c)) depending on the DG pitch $d$. Talbot carpets are shown by color 2D-maps.

An example of such calculations is illustrated in Fig. 3, which shows the fraction of the incident optical intensity ($I_0$) scattered into different DOs ($I_m$) depending on the pitch $d$ of the hybrid mask proposed. The transmittance is defined as the intensity ratio, $T_m = I_m/I_0$, of the corresponding DO, where $m = 0, \pm 1, \pm 2$. The calculation of $T_m$-values is performed in COMSOL through the S-parameters matrix evaluation by including the specific element with the "Port"-type.

In Fig. 3 only five first DOs are plotted because the order number is limited by the condition $|m| < d/\lambda$ that for the considered DG gives $|m| \leq 2$. Since the diffraction grating in question is characterized by the symmetry with respect to the axis perpendicular to the lattice plane and passing through the geometric center of the projection, the transmission $T_{\pm m}$ of positive and negative DOs with the same modulo index will also be the same. Therefore, Fig. 3 shows the total transmittance $T_m$ of each pair of diffraction orders. The simulations are performed for a TE-polarized optical wave normally incident on a dielectric DG with phase steps of the width $w = 240$ nm combined with a shielding Au-mask with $w_m = 210$ nm and $h_m = 50$ nm.

It follows from the analysis of this figure that in the range of values of the mask pitch $\lambda < d < 2\lambda$, the Talbot carpet is formed only by the first three DOs with $m = 0, \pm 1$. According to the

Talbot theory [25], this is the minimum number of diffraction orders which can produce a complete self-image picture. Higher diffraction orders of $m = \pm 2$ occur only at $d > 2\lambda$.

Importantly, in the range of pitch period values 0.45 nm $< d <$ 0.55 nm the fraction of forward scattered optical radiation, which constitutes the zero order of diffraction $T_0$, decreases sharply demonstrating the absolute minimum, $T_0 \approx 10^{-5}$ at $d = 502$ nm (DG filling factor $\kappa \approx 0.48$, and opening $\gamma = 0.58$). This demonstrates that in this case the Talbot diffraction pattern is formed only by the first two orders $T_{\pm 1}$. Just within this region of hybrid mask pitch the periodic diffraction structure loses its longitudinal variation, and Talbot stripes are realized.

It is worth noting that when using only a single phase mask without an amplitude screen, or only a single amplitude mask without phase steps, it is not possible to achieve complete quenching of zero-order diffraction and the occurrence of longitudinally uniform fringes. It is the combination of the amplitude and phase masks that gives the desired effect. This contradicts the results of Ref. [16], where a homogeneous Talbot pattern always appeared for a single $\pi$-shifted phase binary mask under the condition $\kappa = 0.5$. Basically, the reason for this discrepancy lies in the very method of diffraction pattern calculation used in [16], where the scalar approximation to the diffraction theory is used based on the Fourier integrals rather than the rigorous full-wave near-field calculations. Moreover, in our study the phase step of the hybrid mask is considered not as a lumped planar transparence with a $\pi$-phase jump as, e.g., in Ref. [26], but as true 2D-structures with real geometrical parameters. Such diffraction elements redistribute the optical fluxes leaving the ridges and lead to more complicated diffraction pattern formed behind the grating causing an increase in its detail and the appearance of additional intensity maxima (side lobes).

Interestingly to note in Fig. 3 the appearance of a structural resonance in the hybrid mask with the pitch value $d = 700$ nm, which manifests itself as a dip in the total mask transmittance $T$. As noted in [19], such structural resonance characterized by the high spatial localization of the optical field inside the phase ridge and results from the excitation of the Fano resonance in 2D phase DG integrated with an apodizing mask. Fano resonances originate as the constructive interference of coherent multiple scattered waves on the phase steps containing a shielding mask and the optical Mie-type resonance excited in each step.

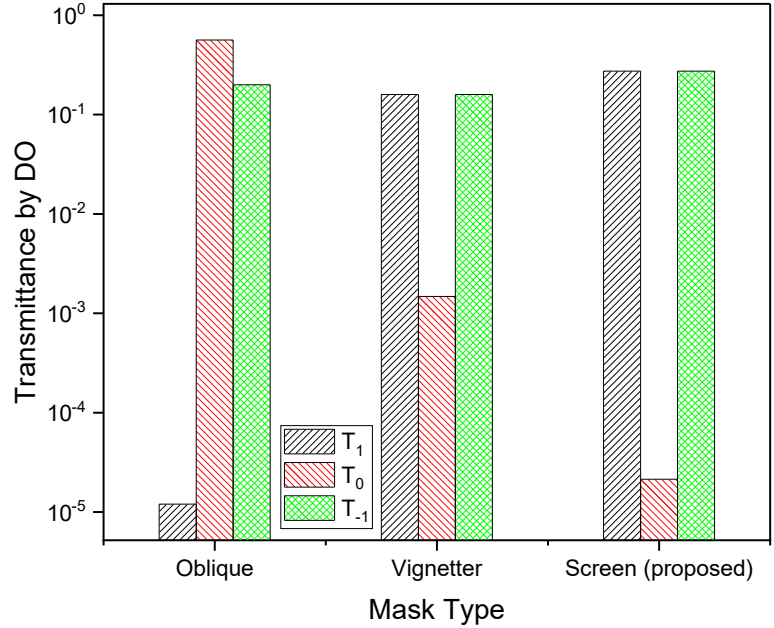

Fig. 4. DO transmittance for different Talbot masks in Figs. 1(a)-(c).

In terms of DO theoretical approach, different ways for Talbot stripes obtaining result in the different ratios of mask transmittance in the diffraction angles considered. This is illustrated in Fig. 4 showing the relative transmittance of the first three diffraction orders for the TL masks presented in Figs. 1(a)-(c) and producing the longitudinally homogeneous Talbot stripes exemplified in Figs. 2(a)-(c). For convenience, different TL schemes are named according to the way of particular DO quenching: "Oblique"- Fig. 1(a), "Vignetter"- Fig. 1(b), "Screen"- Fig. 1(c).

It is clear that the oblique illumination of an ordinary binary phase DG causes a sharp decrease in the intensity of one of the first DO (in this case $T_{-1}$), and the two remaining orders, $T_0$ and $T_1$, form a linear Talbot pattern. On the contrary, the schemes with vignetting and shielding of phase steps by a metal mask suppress zero diffraction order $T_0$, with both first orders $T_{\pm 1}$ having equal intensity due to grating symmetry. Thus, one can conclude that the Talbot stripes in oblique mask illumination are predominantly refractive in nature due to the leading role of close-to-forward scattered waves ($T_0$). In the cases of normal mask exposure shown in Figs. 1(b)-(c), Talbot stripes result from the constructive interference of two waves ($T_{\pm 1}$) of equal intensity propagating at an angle to the normal forming a two-inclined-wave interference pattern.

It is important to compare the quality of the Talbot stripes obtained by different considered TL schemes. Obviously, in practice it is most preferable to have the greatest longitudinal uniformity of fringes intensity with minimum width and maximum transverse optical contrast. Following this reasoning, consider three merits parameters of the diffraction fringe, namely the half-width (FWHM) $W$, longitudinal $C_x$ and transverse $C_y$ optical contrast determined by the difference between maximum $I_{max}$ and minimum $I_{min}$ intensity values in the fringe along the longitudinal and lateral directions according to the Weber definition [27]: $C_{x|y} = 10\log\left[\left(\bar{I}_{max} - \bar{I}_{min}\right)/\bar{I}_{min}\right]$, where the symbols with an upper bar ($\bar{I}$) denote the values considered within two Talbot lengths $z_T$. This definition usually serves as a convenient measure of the difference between the signal value ($I_{max}$) and the intense background ($I_{min}$). The fringe quality parameters are shown in Figs. 5(a,b) for different considered schemes of Talbot stripes acquisition. Note that all values are obtained from the large-scale numerical simulations, and in each case the final parameter value provides the best quality of the Talbot stripes.

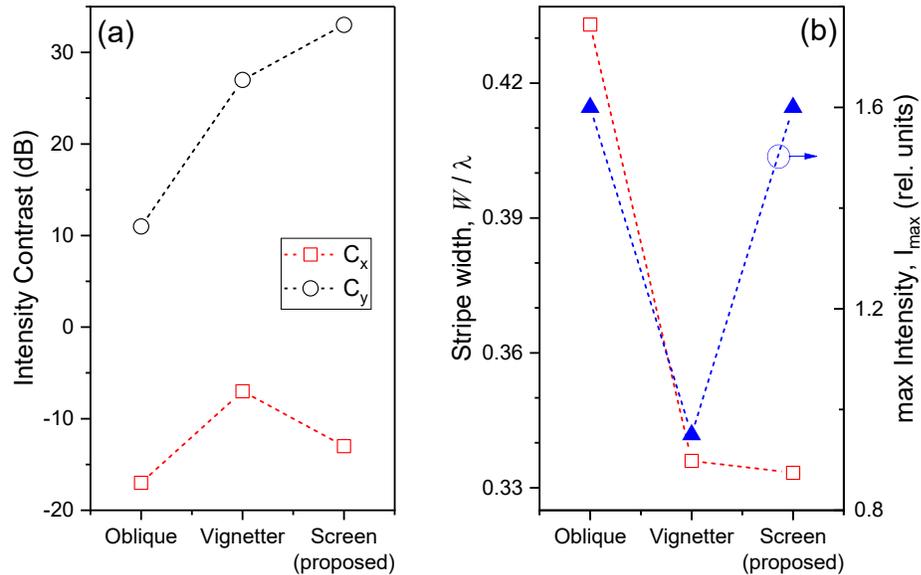

Fig. 5. (a) Longitudinal $C_x$ and lateral $C_y$ optical contrast; (b) FWHM $W$ and maximal intensity $I_{max}$ of the Talbot stripes produced by different masks. The points are connected for visibility.

As follows from this figure, the oblique-illumination scheme produces the greatest longitudinal homogeneity of the diffraction stripes (the lowest $C_x$ value). However, the fringes themselves have the greatest half-width, $W \sim \lambda/2$, and the worst lateral contrast of the Talbot images,

$C_y \approx 11$ dB. The double vignetting mask (Fig. 1(b)) gives uniform stripes with good transverse contrast, $C_y \approx 26$ dB, small half-width, $W \sim \lambda/3$, but with noticeable longitudinal variability, $C_x \approx -6$ dB. In addition, the maximal intensity in the stripes is almost twice as low as in the other cases. Finally, the hybrid shielding mask proposed here, at rather low longitudinal contrast, $C_x \approx -14$ dB, produces Talbot diffraction stripes with the smallest half-width, $W < \lambda/3$, and the best transverse contrast, $C_y \approx 33$ dB that makes possible tight photonic integration for efficient meta-devices.

### b. Talbot stripes robustness and reproducibility

The numerical simulations carried out in a wide range of mask structural parameters allowed to establish with high accuracy the optimal hybrid mask geometry for obtaining a uniform Talbot stripes. However, in the real situation when manufacturing the projection mask, it seems technologically challenging to fabricate exact that mask which was calculated in the theory. Therefore, it is important to understand how sensitive the parameter of fringe longitudinal uniformity is to the changes of the basic mask geometric parameters. Such analysis for the longitudinal contrast is shown in Figs. 6(a-c) upon slightly changing the mask opening ratio $\gamma$, fill factor $\kappa$ and the angle of illumination.

In each graph of Fig. 6 a dashed line is drawn showing the conditional threshold of the contrast, $C_x = -10$ dB, when the Talbot stripes can be considered inhomogeneous. As seen, the range of possible values for each of the considered structural parameters is quite wide and exceeds 6%. The greatest sensitivity the longitudinal contrast exhibits at the variation of the phase steps width $w$. The deviation from the normal incidence of optical radiation illuminating the mask is not critical up to the inclination angles $\alpha \approx 5°$ maintaining the longitudinal and transverse contrasts of the diffraction fringes at an acceptable level.

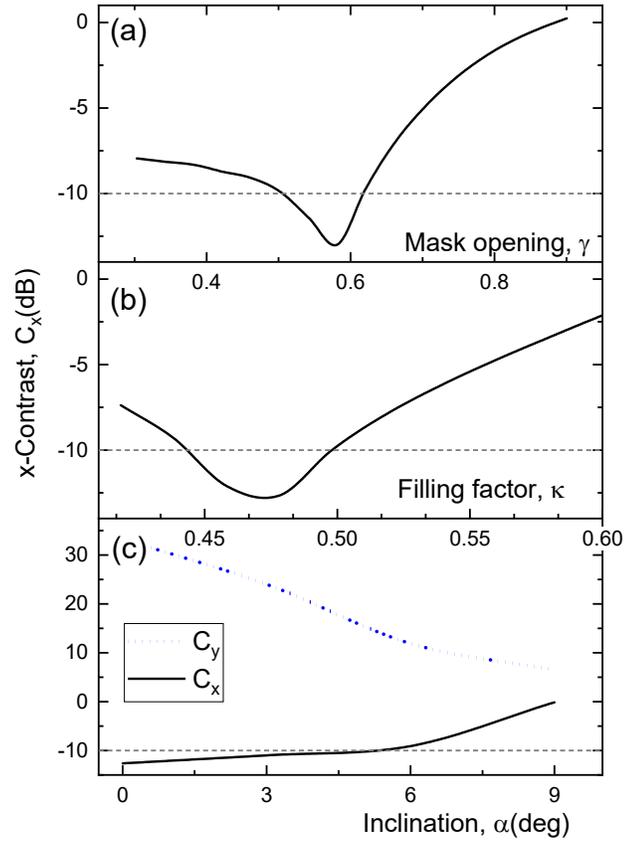

Fig. 6. Talbot stripe tolerance to hybrid mask parameters: Longitudinal contrast $C_x$ versus (a) mask opening $\gamma$, (b) phase step filling factor $\kappa$, and (c) illumination angle $\alpha$. Additionally, the transverse optical contrast $C_y$ is shown in (c).

Figure 7(a) demonstrates the reproducibility of the Talbot longitudinal stripes when changing the structural design of the hybrid mask and the polarization of the illuminating optical radiation. Here, the transmittance for zero-order diffraction $T_0$ is plotted as a function of the mask pitch $d$. In the first case (green dashed line), we replace the material of the dielectric grating with titanium dioxide ($TiO_2$), which has much higher refractive index in the UV spectrum region than silica, $n = 2.419$. This requires reducing the height of the phase steps to the value of $h = 140$ nm for maintaining the phase shift of the optical wave per $\pi$-radian. As can be seen from Fig. 7(a), for such a Talbot mask the effective quenching of the zero-order diffraction occurs at the same value of the phase mask period, $d = 502$ nm, as in the case of the silica matrix. Using a material with stronger refraction, here can be more promising in terms of obtaining quasi-flat (quasi 1D) projection masks.

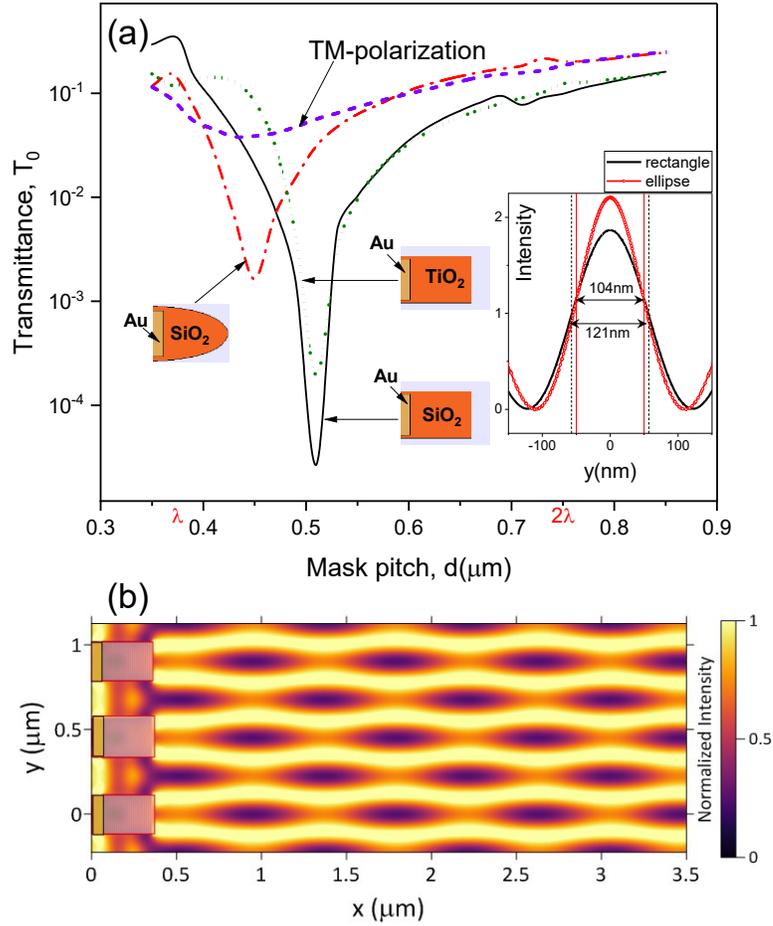

Fig. 7. (a) Transmittance of diffraction zero-order $T_0$ for different hybrid masks and light polarization (sown by pictograms); (b) Talbot stripes with TM-wave.

Next calculations we perform again for a silica mask but with elliptical-section ridges (red dashed line). As seen, the required longitudinally uniform Talbot carpet appears at higher phase mask fill factor $\kappa = 0.56$ versus 0.48 for a rectangular mask. At the same time, due to incomplete suppression of zero-order diffraction for this type of mask, we could not achieve a longitudinal contrast of the diffraction stripes better than $C_x = -8.5$ dB. Meanwhile, the advantage of ellipsoidal phase masks is the higher peak intensity and smaller half-width of the diffraction fringes, which reaches the value $W \approx 104$ nm ($\lambda/3.6$) (see inset to Fig. 7(a)). Such resolution is achievable for Talbot lithography based on solid immersion technology [21].

Finally, the third series of numerical experiments is performed with a different polarization of the optical wave. In contrast to the previously used *s*-polarized wave (TE), a *p*-polarized wave

(TM) with nonzero electric field vectors $E_x$, $E_y$ and a zero $z$-component is now incident onto the silica rectangular mask (purple dotted curve in Fig. 7(a)). It turned out that in this case, due to the presence of a field component orthogonal to the DG grooves, the hybrid mask exhibits features of a plasmonic structure when a surface-propagating mode is excited in the metal strips. The coupling of the surface plasmon and the optical radiation passing through the phase steps results in the appearance of an optical vortex in the center of the ridge [19], a local reversal of the optical energy flux, and a stronger refraction of the direct radiation, which now cannot be effectively suppressed by tailoring the geometric structure of the mask. As a result, the field diffraction pattern after the optical mask takes the form of Talbot stripes with wavy boundaries whose self-repeating period is equal to the Talbot length, $z_T = 840$ nm (Fig. 7(b)). The longitudinal contrast of the diffraction fringe calculated along its median is $C_x = -12$ dB with weak transverse separation, $C_y = 0.9$ dB.

### c. Talbot stripes from hybrid 2D-mask

Above we have discussed various schemes for obtaining Talbot diffraction fringes from 1D projection masks. It is of interest to consider the possibility of creating a Talbot 2D-carpet formed by diffraction fringes with minimal longitudinal intensity variability. Obviously, for this purpose it is also necessary to use a 2D-mask, which will consist of a regular matrix of phase steps, and a metal amplitude grating combined with it. The building block of the proposed periodic 2D-mask is shown in Fig. 8(a) and consists of a rectangular block on the lower base of which a phase step together with a rectangular gold mask are formed. Periodic structure of the mask is provided by setting the periodic boundary conditions (denoted as "Periodic BC" in the figure) on the side faces of the model block following the Floquet-Bloch relations for optical field vectors. The lower face of the building block is illuminated by a plane optical wave with $\lambda = 375$ nm in the direction of the wave vector **k**, and on the opposite face the free propagation conditions in the form of perfectly-matched layers (PML) are set. The computational volume of the model is discretized by a spatial 3D-grid with an adaptive step, which size decreases in regions with large values of the complex medium permittivity. The minimum mesh size is 0.2 nm and the maximum is 10 nm. Numerical simulation of the near-field scattering of electromagnetic wave on a 2D-mask is performed using the FDTD technique implemented in the commercial package *Lumerical FDTD* (ver. 8.19) to reduce time and computer resources.

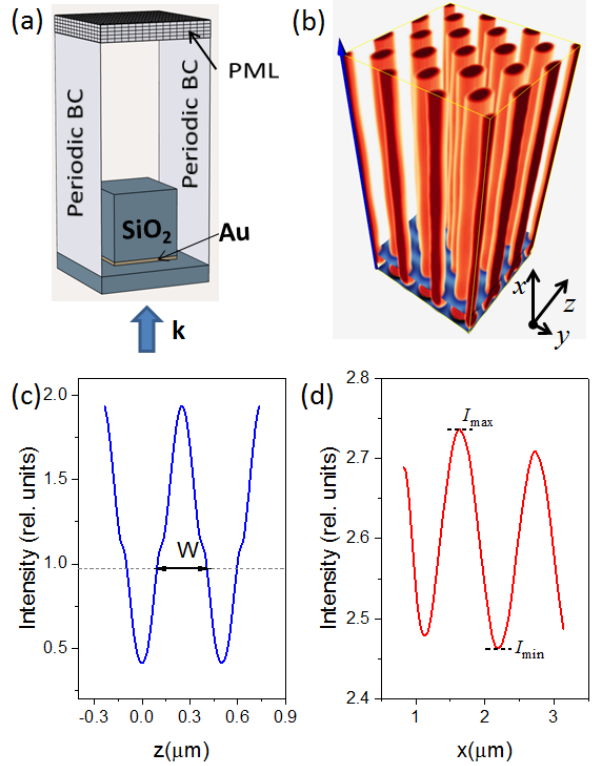

Fig. 8. (a) The building block of the hybrid Talbot 2D-mask; (b) Volume rendering of 2D-array of Talbot stripes; (c) Lateral and (d) longitudinal intensity profiles of Talbot fringe.

The spatial structure of the optical intensity formed by the hybrid 2D-mask in air is shown in Fig. 8(b) as a volumetric 3D-image of an ensemble of Talbot stripes. To obtain this diffraction pattern, the mask parameters are chosen as follows: $d$ = 500 nm, $w$ = 250 nm, $h$ = 530 nm, $h_m$(Au) = 60 nm, $w_m$(Au) = $w$. Due to the fact that the incident wave is linearly polarized along the $z$-axis, each diffraction stripe has an elliptical cross-section. In this case, the minimum half-width of the Talbot stripe (across the wave polarization) is on the order of 170 nm. Meanwhile, the maximum half-width is also subwavelength, $W \approx$ 257 nm, as seen in Fig. 8(c). The longitudinal optical contrast of the diffraction fringes is sufficiently low and according to Fig. 8(d) is $C_x$ = -10.8 dB. Worthwhile noting, the obtained regular fringe structure of Talbot stripes may also be interesting for creating a "Talbot crystal" (a photonic crystal based on Talbot stripes) similar to that described earlier in [28].

## IV. Conclusions

To conclude, for the application in the Talbot photolithography we propose and theoretically study a new way for creating a periodic longitudinally-uniform Talbot carpet by a hybrid metal-dielectric projection mask of wavelength-scale when illuminated by a linearly polarized monochromatic optical radiation. The hybrid mask consists of a binary phase diffraction grating and an amplitude (Au) screen embedded in the base of each phase step. We consider in detail the conditions for obtaining a specific diffraction pattern behind the mask exhibiting an ensemble of equally spaced longitudinally uniform diffraction fringes having subwavelength half-width at minimal intensity variation along the direction of light propagation. Using the numerical FEM & FDTD simulations of the near-field electromagnetic wave diffraction, we determine the particular values of structural mask parameters when the striped Talbot carpet with high longitudinal homogeneity and maximum transverse optical contrast can be obtained. The physical analysis is carried out by inspecting the contribution of different diffraction orders of light scattered at the hybrid mask to the total Fresnel self-images pattern. It turned out that the formation of longitudinally homogeneous diffraction stripes after a Talbot transparence is only possible as a result of the two-wave interference, i.e. the interference of only two diffraction orders. In this connection, a metal screen of the proposed hybrid mask provides shielding the direct-scattered light coming in the zero-order diffraction, thus enabling a sharp increase in the contributions of optical waves propagating in the spatial angles of the negative and positive first diffraction orders. The constructive wave interference from these diffraction orders produces the desired Talbot stripes.

We compare the optical quality of the Talbot diffraction fringes obtained with the proposed hybrid mask and with other known Talbot lithography schemes (oblique illumination, vignetting combined mask) and demonstrate the obvious advantages of our technical design. In particular, by properly tailoring the fused silica hybrid mask it is possible to obtain the Talbot stripes with a minimal longitudinal optical contrast (-14 dB) and a high transverse contrast (>30 dB) at sub-wavelength spatial resolution (< $\lambda/3$). Worth noting, in this paper the very specifics of the proposed bi-component diffraction mask fabrication is not discussed. However, similarly structured reflecting hybrid diffraction gratings are already commercially available aimed to broader spectral and angular bandwidths of diffracted light.

**Funding**
The work is partially supported by the Ministry of Science and Higher Education of the Russian Federation and by TPU development program.

**Conflict of interest**
The authors declare no conflicts of interest.